\documentclass[useAMS,usenatbib]{mn2e}
\usepackage[dvips]{graphicx}
\usepackage{longtable}

\title[Type Ia Supernova SN 2003hx]{Photometric and Spectroscopic study of a highly reddened type Ia supernova SN 2003hx in NGC 2076}

\author[Kuntal Misra et al.]
{Kuntal Misra$^{1,2}$, D. K. Sahu$^3$, G. C. Anupama$^4$ and Kavita Pandey$^5$ \\
1.Aryabhatta Research Institute of Observational Sciences, Manora Peak, Nainital 263 129, India\\
2.Inter University Center for Astronomy and Astrophysics, Pune 411 007, India\\
3.Center for Research and Education in Science \& Technology, Hosakote, Bangalore 562 114, India\\
4.Indian Institute of Astrophysics, Bangalore 560 034, India\\
5. Department of Physics, Kumaun University, Nainital 263 002, India\\
(E-mail:kuntal{@}aries.ernet.in, dks{@}crest.ernet.in, gca{@}iiap.res.in, kavitaphysics{@}yahoo.co.uk)}

\begin{document}

\date{Accepted.....; Received .....}

\pagerange{\pageref{firstpage}--\pageref{lastpage}} \pubyear{}

\maketitle

\label{firstpage}

\begin{abstract}
We present $UBVRI$ CCD photometry and optical spectra of the type Ia supernova SN 2003hx
which appeared in the galaxy NGC 2076, obtained till $\sim$ 146 days after the epoch of
$B$ band maximum. The supernova reached at maximum brightness in $B$ band on JD 245 2893 $\pm$ 1.0
with an apparent magnitude of 14.92 $\pm$ 0.01 mag which was estimated by making template fits
to the light curves. SN 2003hx is an example of a highly reddened supernova 
with $E(B-V)$ = 0.56 $\pm$ 0.23. We estimate $R_v$ = 1.97 $\pm$ 0.54 
which indicates the small size of dust particles as compared to their galactic counterparts.
The luminosity decline rate is $\Delta m_{15}(B)$ = 1.17 $\pm$ 0.12
mag and the absolute $B$ band magnitude obtained from the luminosity versus decline rate
relation (Phillips et al. 1999) is $M^B_{max}$ = -19.20 $\pm$ 0.18 mag. The peak
bolometric luminosity indicates that $\sim$ 0.66 $M_\odot$ mass of $^{56}$ Ni was ejected by the
supernova. The spectral evolution indicates the supernova to be a normal type Ia event.
\end{abstract}


\section{INTRODUCTION}
\label{introduction}
Type Ia supernovae, being one of the most luminous stellar outburst, form a fairly
homogeneous class of objects and are considered as standard candles for determining
extragalactic distances and cosmological parameters. Type Ia supernovae are produced
by thermonuclear explosions of white dwarfs (Hoyle \& Fowler 1960) primarily composed
of Carbon and Oxygen nuclei. The probable explosion scenario involves a binary
system in which a white dwarf accretes matter from its companion star until it reaches
the Chandrasekhar mass limit of 1.4 M$_{\odot}$. Though type Ia supernovae show
homogeneity in both their photometric as well as spectroscopic properties
(H\"{o}flich et al. 1996) but many SNe show significant deviations. Li et al. (2001)
studied a sample of type Ia supernovae and conclude that $\sim$ 64 percent of them
belong to the normal group, almost 20 percent belong to the over luminous group of events
such as SN 1991T whereas about 16 percent belong to the sub luminous type such as
SN 1991bg.

The last few decades have witnessed a large number of diverse data sets
for SNe Ia. Nevertheless, type Ia supernovae seem to follow a few common patterns.
One of these is the correlation between the peak luminosity and the linear decline rate
(Phillips 1993). The color evolution, spectral appearance and the host galaxy morphology
are the other correlations. The peak absolute magnitude of type Ia SNe is correlated with
the Hubble type of the parent galaxy. SNe Ia hosted by elliptical galaxies are comparatively
fainter than SNe Ia in spirals (Della Valle \& Panagia 1992, Howell 2001). Even amongst the
normal group of type Ia SNe significant photometric and spectroscopic uncertainties exist.
Nugent et al. (1995) saw that the spectral variations in type Ia SNe correlate with the
expansion velocity, the effective temperature and the peak luminosity. Thus, it is not
sufficient to describe SNe Ia by a single parameter such as the early light curve decline
rather the diversity noticed is multi-dimensional (Hatano et al. 2000, Benetti et al. 2004).
It is therefore, necessary to study the photometric and spectroscopic evolution of
individual type Ia SNe.

The integrated flux in optical bands provides a meaningful estimate of the bolometric luminosity
which is directly related to the amount of radioactive
$^{56}$Ni synthesised and ejected in the explosion (Arnett 1982, H\"{o}flich et al. 1996,
Pinto \& Eastman 2000a) which can later be used to test various explosion models of type Ia
supernovae.

We present in this paper the optical photometric and spectroscopic observations of
highly reddened type Ia supernova SN 2003hx.
SN 2003hx was discovered on unfiltered KAIT images on 2003 September 12.5 UT (magnitude 14.3)
and 13.5 UT (magnitude 14.4) by Burket, Papenkova \& Li (2003). A KAIT image of the same
region taken on 2003 March 7.2 UT shows nothing at the location of the supernova to a
limiting magnitude of $\sim$ 18.5. SN 2003hx is located at
$\alpha$ = $05\rm^{h} 46\rm^{m} 46\rm^{s}.97$,
$\delta$ = $-16^{\circ} 47\arcmin 00\arcsec.6$ (J2000)
which is 5$\arcsec$.2 West and 2$\arcsec$.6 South of the nucleus of the galaxy
NGC 2076. A spectrum of SN 2003hx taken on 2003 September 13.78 UT at the Australian
National University (ANU) 2.3-m telescope shows it to be a type Ia supernova around maximum
light (Salvo, Norris \& Schmidt, 2003). The Si II 635.5 nm line gives an expansion velocity
of $\sim$ 12000 km/sec if the NED recession velocity for the host is adopted as 2142 km/sec
(Salvo, Norris \& Schmidt, 2003). It was confirmed to be a type Ia supernova around 10 days
past maximum light with the spectropolarimetric observations using the ESO very large telescope
on 2003 September 15.4 UT (Wang \& Baade, 2003). The interstellar Na I D line has an
equivalent width of 4.86 \AA which indicates significant dust extinction. The observed degree
of polarization is $\sim$ 2 percent (Wang \& Baade, 2003). If this polarization is due to the
dust in the host galaxy, then it implies that the dust particles are significantly smaller
in size than their Galactic counterparts. These observations yield a value of 2.2 for the
ratio of total to selective extinction, $R_v = A_v/E(B-V)$ (Wang \& Baade, 2003).

We have carried out the optical photometric and spectroscopic observations of the
type Ia supernova SN 2003hx. A brief description of the observations and data analysis
is given in section \ref{observations}, whereas the development of the light curves and color curves
are presented in section \ref{light_curves}. Section \ref{reddening} discusses about the reddening
estimate. The description about the absolute magnitude, the
bolometric luminosity and the estimation of $^{56}$Ni ejected are discussed in section \ref{bolometric}.
Spectral evolution has been studied with a comparison to other type Ia supernovae in
section \ref{spectra}. Conclusion forms section \ref{conclusion} of the paper.

\section{OBSERVATIONS AND DATA REDUCTION}
\label{observations}
The observations of SN 2003hx were carried out with the 2-m Himalayan Chandra Telescope (HCT) at
Indian Astronomical Observatory (IAO), Hanle during 2003 September 18 to 2004 February 03 which was
six days after the discovery on 2003 September 12. The Himalayan Faint Object Spectrograph Camera
(HFOSC) equipped with the SITe 2 K $\times$ 4 K pixel CCD was used. The central 2 K $\times$ 2 K
region was used for imaging covering a field of 10 \arcmin $\times$ 10 \arcmin on the sky
corresponding to a plate scale 0.296 arcsec pixel $^{-1}$. The gain and read out noise of the
CCD camera are 1.22 e$^-$/ADU and 4.87 e$^-$ respectively.
\subsection{Photometry}
\begin{figure}
\centering
\includegraphics[width=\columnwidth]{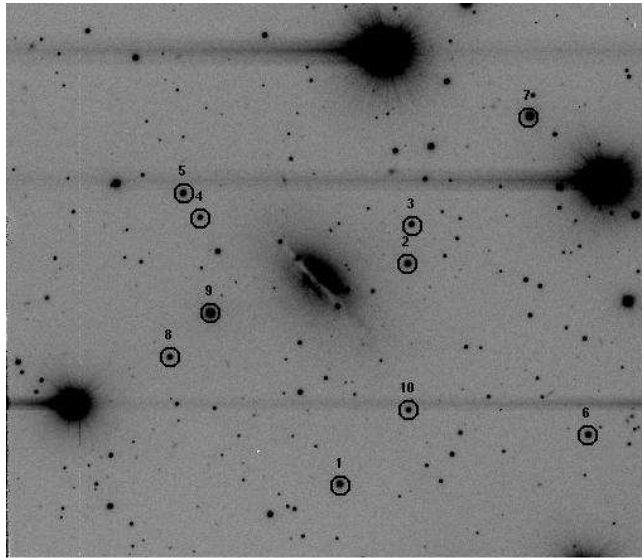}
\caption{A 10$\arcmin$ $\times$ 10$\arcmin$ field of SN 2003hx. The 10 secondary stars used for
calibration are marked. North is up and East is to the left.}
\label{chart_calib_stars}
\end{figure}

The broad band $UBVRI$ photometric observations of SN 2003hx were carried out at 16 epochs during
18 September 2003 to 03 February 2003. All the images were bias subtracted, flat fielded and
cosmic ray removed in the standard fashion using various tasks in IRAF. Landolt (1992) standard
region PG 0231+051 was imaged along with the supernova field in $UBVRI$ filters on 26 October 2005
under good photometric sky conditions. The values of atmospheric extinction on the night of
26/27 October 2005 determined from the observations of PG 0231+051 bright stars are
0.30 $\pm$ 0.01, 0.20 $\pm$ 0.009, 0.12 $\pm$ 0.007, 0.08 $\pm$ 0.004 and 0.04 $\pm$ 0.003
in $U, B, V, R$ and $I$ filters respectively. The observations of PG 0231+051 were used to
generate secondary standards in the supernova field. The $UBVRI$ magnitudes of 10 secondary
stars using the transformation equations obtained are listed in Table \ref{secondary_standards}
and are marked in Figure \ref{chart_calib_stars}. These magnitudes were used to calibrate the
data obtained on other nights. The sequence photometry of SN 2003hx field was carried out
by the American Association of Variable Star Observers (AAVSO). We compared our field
calibration with that of AAVSO and found it to be consistent. Thus, our calibration
of SN 2003hx field is secure.

\begin{figure}
\centering
\includegraphics[width=\columnwidth]{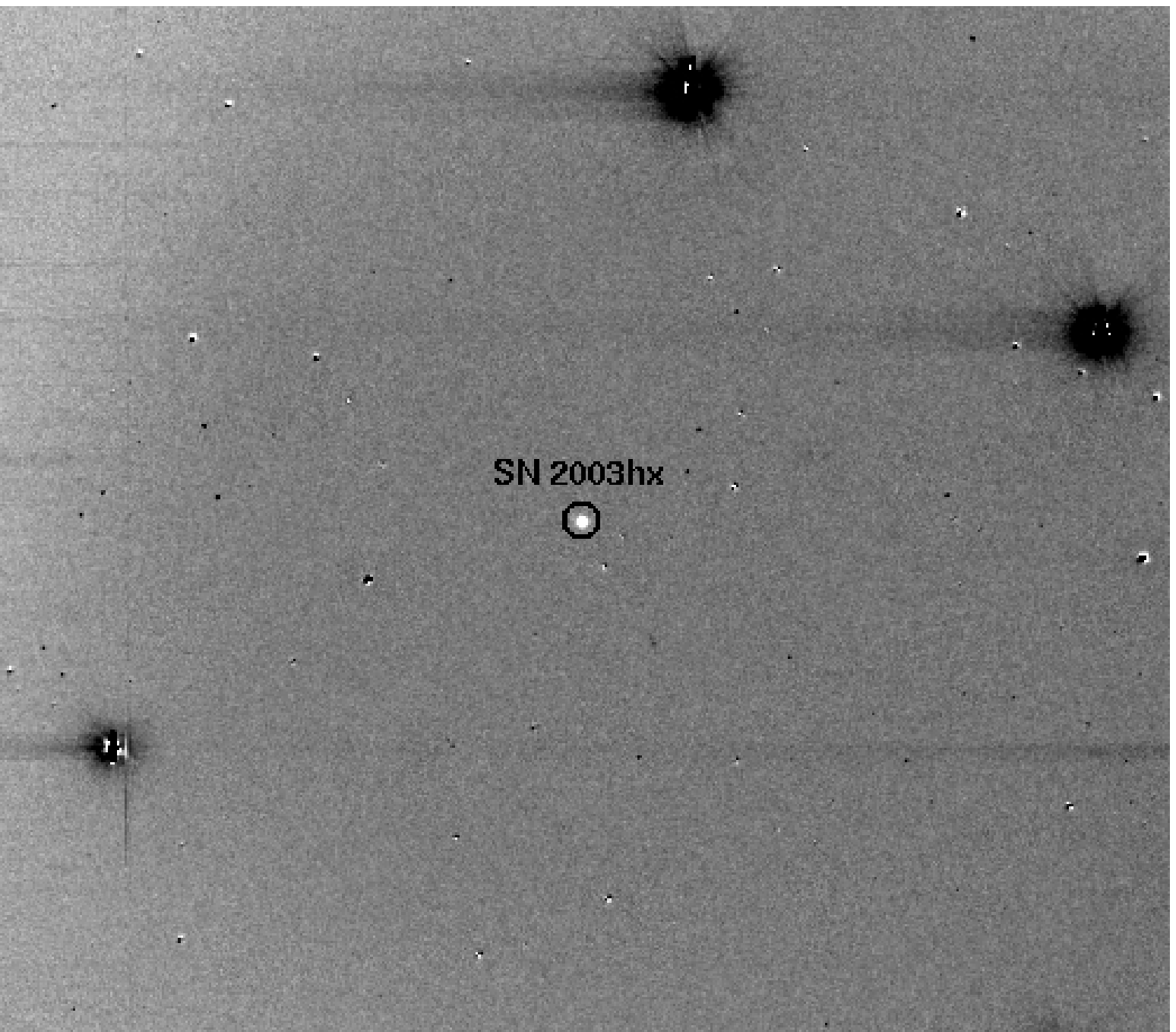}
\caption{Supernova SN 2003hx after the template subtraction.}
\label{sn2003hx}
\end{figure}

We performed aperture photometry on the local standards using an aperture of 3 - 4 times the
$FWHM$ of the seeing profile that was determined on the basis of an aperture growth curve. Accurate
estimate of sky and its subtraction plays a crucial role in the photometry when the object
falls on the varying background.
The supernova was just next to the nucleus of the galaxy, we therefore had to perform template subtraction
to get a better estimate of the underlying background.
The template observations were taken with the same instrumental setup, on 26 October 2005,
nearly two years after the supernova discovery, when the supernova had faded. These templates were
subtracted from the images of SN obtained during 2003 - 2004. One such subtracted image, showing
only the supernova SN 2003hx, is shown in Figure \ref{sn2003hx}. When the seeing was not very good,
the template subtraction was not perfect and showed signature of the galaxy. Here, the supernova
magnitudes were determined by using the profile-fitting method with a fitting radius equal to that
of the $FWHM$ of the seeing profile. The difference between the aperture and profile-fitting
magnitudes was obtained using standards and was applied to the supernova magnitudes. Thus, the
supernova magnitudes were obtained by differentially calibrating with respect to the secondary
standards listed in Table \ref{secondary_standards}. The supernova magnitudes derived in this way
are given in Table \ref{photometric_observations}.
%
\begin{table*}
\caption{Magnitudes for the sequence of secondary standard stars in the field of SN 2003hx. The stars
are identified in Figure \ref{chart_calib_stars}.}
\medskip
\begin{center}
\begin{tabular}{cccccc} \hline\hline
&&&&&\\
ID & $U$ & $B$ & $V$ & $R$ & $I$\\ \hline
1  & 17.17 $\pm$ 0.07 & 16.22 $\pm$ 0.01 & 15.20 $\pm$ 0.01 & 14.65 $\pm$ 0.02 & 14.12 $\pm$ 0.01\\
2  & 15.43 $\pm$ 0.07 & 15.60 $\pm$ 0.01 & 15.09 $\pm$ 0.01 & 14.74 $\pm$ 0.02 & 14.36 $\pm$ 0.01\\
3  & 16.58 $\pm$ 0.07 & 16.51 $\pm$ 0.01 & 15.88 $\pm$ 0.01 & 15.48 $\pm$ 0.02 & 15.07 $\pm$ 0.01\\
4  & 16.52 $\pm$ 0.07 & 16.74 $\pm$ 0.01 & 16.29 $\pm$ 0.01 & 15.96 $\pm$ 0.02 & 15.63 $\pm$ 0.01\\
5  & 16.09 $\pm$ 0.07 & 15.95 $\pm$ 0.01 & 15.32 $\pm$ 0.01 & 14.91 $\pm$ 0.02 & 14.51 $\pm$ 0.01\\
6  & 16.29 $\pm$ 0.07 & 16.13 $\pm$ 0.01 & 15.47 $\pm$ 0.01 & 15.08 $\pm$ 0.02 & 14.63 $\pm$ 0.01\\
7  & 14.66 $\pm$ 0.07 & 14.65 $\pm$ 0.01 & 14.14 $\pm$ 0.01 & 13.81 $\pm$ 0.02 & 13.48 $\pm$ 0.01\\
8  & 16.58 $\pm$ 0.07 & 16.10 $\pm$ 0.01 & 15.81 $\pm$ 0.01 & 15.42 $\pm$ 0.02 & 15.02 $\pm$ 0.01\\
9  & 15.10 $\pm$ 0.07 & 14.70 $\pm$ 0.01 & 13.96 $\pm$ 0.01 & 13.52 $\pm$ 0.02 & 13.13 $\pm$ 0.01\\
10 & 17.29 $\pm$ 0.07 & 16.92 $\pm$ 0.01 & 16.24 $\pm$ 0.01 & 15.84 $\pm$ 0.02 & 15.42 $\pm$ 0.01\\
&&&&&\\
\hline
\end{tabular}
\label{secondary_standards}
\end{center}
\end{table*}
%
%
\begin{table*}
\caption{Photometric observations of SN 2003hx}
\medskip
\begin{center}
\begin{tabular}{cccccccc} \hline \hline
&&&&&&&\\
Date & JD & Phase$^*$ & $U$ & $B$ & $V$ & $R$ & $I$\\
&2,400,000+& (days)&&&&& \\ \hline
&&&&&&&\\
18/09/03 & 52901.4769 &  8.47 & 16.20 0.03 & 15.07 0.03 & 15.07 0.03 & 14.98 0.03 & 14.89 0.02\\
20/09/03 & 52903.4866 & 10.46 &            & 15.24 0.03 & 15.19 0.05 & 14.93 0.04 &           \\
21/09/03 & 52904.4847 & 11.48 &            & 15.34 0.04 & 15.22 0.05 & 14.94 0.05 & 14.73 0.04\\
22/09/03 & 52905.4805 & 12.48 &            & 15.46 0.03 & 15.12 0.03 & 14.89 0.02 & 14.53 0.02\\
27/09/03 & 52910.4865 & 17.48 &            & 16.09 0.22 & 15.46 0.05 & 14.87 0.04 & 14.40 0.03\\
28/09/03 & 52911.4870 & 18.48 &            & 16.23 0.06 & 15.63 0.05 & 14.90 0.04 & 14.30 0.06\\
06/10/03 & 52919.4786 & 26.47 & 17.11 0.21 &            & 16.17 0.02 & 15.51 0.01 & 15.17 0.03\\
18/10/03 & 52931.4155 & 38.41 &            & 18.06 0.03 & 16.65 0.03 & 16.34 0.02 & 15.99 0.05\\
27/10/03 & 52940.3899 & 47.38 &            & 18.23 0.02 & 16.93 0.03 & 16.68 0.03 & 16.44 0.04\\
01/11/03 & 52945.4695 & 52.46 &            & 18.28 0.01 & 17.09 0.02 & 16.81 0.04 & 16.64 0.05\\
02/11/03 & 52946.4872 & 53.48 &            &            & 17.12 0.04 & 16.85 0.03 & 16.71 0.06\\
07/11/03 & 52951.2987 & 58.29 & 18.07 0.10 & 18.30 0.03 & 17.18 0.02 &            & 16.73 0.04\\
18/11/03 & 52962.4231 & 69.42 &            &            & 17.45 0.02 & 17.27 0.02 &           \\
04/01/04 & 53009.2533 &116.25 &            &            & 18.56 0.04 & 18.52 0.06 & 18.12 0.11\\
27/01/04 & 53031.1773 &138.17 &            & 19.41 0.05 & 18.85 0.08 &            & 18.17 0.08\\
03/02/04 & 53039.1489 &146.14 &            &            & 19.24 0.06 & 19.23 0.10 &           \\
&&&&&&&\\
\hline
\end{tabular}
\label{photometric_observations}
\end{center}
* Relative to the epoch of B maximum (this work) JD = 2452 893.0
\end{table*}
\subsection{Spectroscopy}
Spectroscopic observations of SN 2003hx were carried out on four nights starting 19 September 2003 which
was 7 days after the discovery on 12 September 2003. The log of spectroscopic observations is given in
Table \ref{spectroscopic_observations}. All the spectra were obtained at a resolution of $\sim$ 7 \AA \ in
the wavelength range 3300--6000 \AA, 3500--7000 \AA, 5200--9200 \AA and 5200--10300 \AA.
Spectroscopic  data were reduced using
the standard routines within IRAF. The data were bias corrected, flat-fielded and the one
dimensional spectra were extracted using the optimal extraction method.  FeAr and FeNe sources were
used for wavelength calibration.
The wavelength calibrated spectra were corrected for instrumental response using spectra of
spectrophotometric standards observed on the same night and brought to the same flux scale. The
final spectrum on a relative flux scale were obtained by combining the flux calibrated spectra in the
two different regions scaled to a weighted mean.
%
%
\begin{table}
\caption{Spectroscopic observations of SN 2003hx}
\medskip
\begin{center}
\begin{tabular}{cccc} \hline \hline
Date & JD & Phase$^*$ & Range \\
&2,400,000+& (days)& \AA \\ \hline
19/09/03& 2452902.4769 &  9.47 & 3300-6000; 5200-10300\\
22/09/03& 2452905.4805 & 12.48 & 3500-7000; 5200-9200\\
24/09/03& 2452907.1805 & 14.18 & 3500-7000; 5200-9200\\
02/11/03& 2452946.4872 & 53.48 & 3500-7000; 5200-9200\\
 \hline
\end{tabular}
\label{spectroscopic_observations}
\end{center}
* Relative to the epoch of B maximum JD = 2452 893.0
\end{table}
%

\section{Light Curves and Color Curves}
\label{light_curves}
In this section we study the multi band light curve and color curve evolution of SN 2003hx and
compare it with other type Ia supernovae.
\subsection{$UBVRI$ light curves}
\begin{figure}
\centering
\includegraphics[width=\columnwidth]{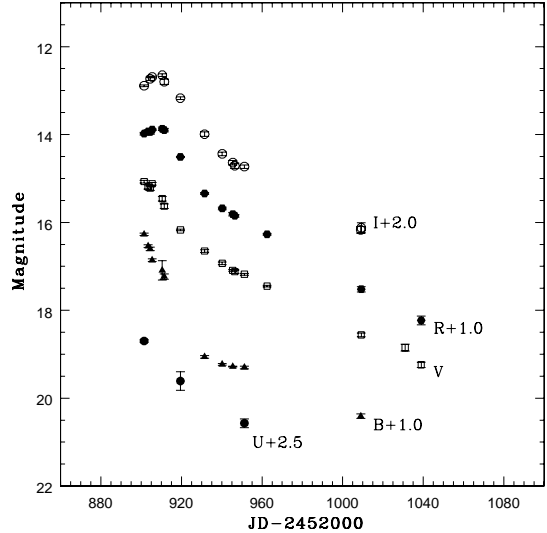}
\caption{$UBVRI$ light curve of SN 2003hx. The light curves are offset by a constant value
on the magnitude scale as indicated in the plot.}
\label{sn2003hx_lc}
\end{figure}

\begin{figure}
\centering
\includegraphics[width=\columnwidth]{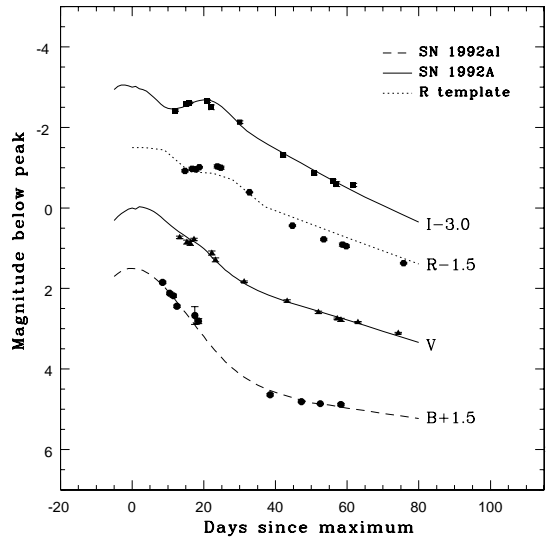}
\caption{$BVRI$ light curves of SN 2003hx with the best fitting templates of Hamuy et al. (1996)
in the $BVI$ bands and Schlegel (1995) in the $R$ band. B band light curve fits best with
SN 1992al template (dashed line: $\Delta m_{15}(B)$ = 1.11) whereas the V and I light curve
fit best with SN 1992A template (solid line). The R band template is shown by dotted lines.
The light curves are offset by a constant value on the magnitude scale as indicated in the plot
for clarity.}
\label{sn2003hx_lc_template}
\end{figure}

We present the $UBVRI$ light curves of SN 2003hx in Figure \ref{sn2003hx_lc}. Since our
observations started $\sim$ 6 days after the discovery, we do not have observations around
the peak brightness. The peak brightness and the JD corresponding to the peak brightness in
different bands was estimated by making template fits to the observations. Our observations
span a period of $\sim$ 150 days.
The frequency distribution of our data is $N(U, B, V, R, I)$ = (3, 11, 16, 14, 13).
In order to determine peak brightness in different bands we have adopted the template
fitting method. We attempted to fit the different template sets in $BVI$ bands given by
Hamuy et al. (1996). We adopted a $\chi ^2$ minimizing technique which solved simultaneously
for the peak magnitude and the peak time in different bands. We see that the $B$ band fits
best with the template of SN 1992al whereas the lowest value of $\chi ^2$ is attained
in the case of SN 1992A for $V$ and $I$ bands. Since Hamuy et al. (1996) does not present
the $R$ band template, we have taken the $R$ band template for type Ia supernova from
Schlegel (1995). The $R$ band template was similarly fit to the observations to determine the
peak magnitude and the peak time in $R$ band. Figure \ref{sn2003hx_lc_template} shows the
light curves in $BVRI$ bands including the template fits to the data.

The main parameters of SN 2003hx as estimated from template fits are listed in Table \ref{parameters}.
Leibundgut (1988) showed that for a normal type Ia supernova, a 2 day difference is seen between the
time of maxima in the $B$ and $V$ bands. We see from the template fits that SN 2003hx reached
maximum brightness in the $V$, $R$ and $I$ bands earlier than in the $B$ band. An excellent match
of the SN 1992al template with that of SN 2003hx indicates that the peak in the $B$ band occurred
at JD 245 2893 $\pm$ 1.0. These peak magnitudes obtained using template fitting are further used
to calculate the peak luminosity in section \ref{bolometric}.

In Figure \ref{sn2003hx_lc_template} we see a pronounced secondary maximum for SN 2003hx in the $I$ band.
This secondary maximum was seen $\sim$ 21 days after the $B$ maximum and was 0.32 magnitude fainter
than the first maximum. The magnitude of the secondary $I$ maxima is listed in Table \ref{parameters}.
$R$ band also shows a noticeable rise similar to the $I$ band at similar epochs. This secondary
maxima in the $I$ band is a remarkable feature of type Ia SNe and becomes more pronounced in the
near-IR band. Such behaviour has been seen for many type Ia SNe. Elias et al. (1981) and Pinto \&
Eastman (2000b) pointed out that this secondary maxima is due to a temporary increase in absorption,
which reduces with the fall in the degree of ionization several weeks after maximum light.

$\Delta m_{15}(B)$, the number of magnitudes in the $B$ band by which the SN declines in the first
15 days after maximum, is a characteristic feature of the type Ia SNe. The fitted template of SN 1992al
has a $\Delta m_{15}(B)$ = 1.11. We calculate $\Delta m_{15}(B)$ = 1.17 $\pm$ 0.12 by taking the $B$
band peak magnitude obtained by the template fit and the observed $B$ band magnitude $\sim$ 15 days
after the $B$ maximum. The average decline rate in different bands is also estimated, and listed in
Table \ref{parameters}, from our observations using a time baseline of 20 days.

\subsection{Color Curves}
\begin{figure}
\centering
\includegraphics[height=8.0cm,width=8.0cm]{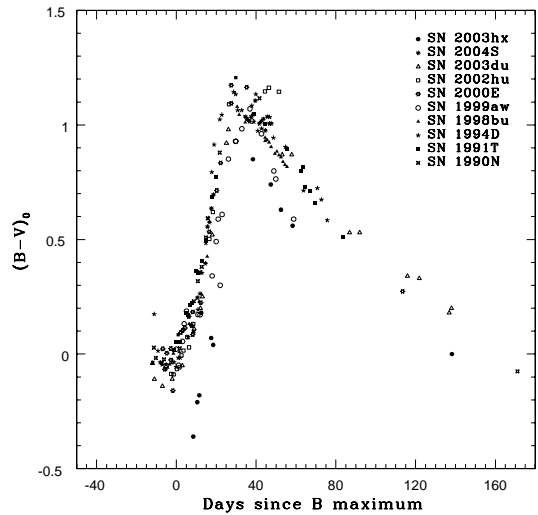}
\caption{The intrinsic $B-V$ color evolution of SN 2003hx compared with other supernovae of
type Ia. The adopted reddening value for each supernova is mentioned in the text.}
\label{bv_color}
\end{figure}

\begin{figure}
\centering
\includegraphics[height=8.0cm,width=8.0cm]{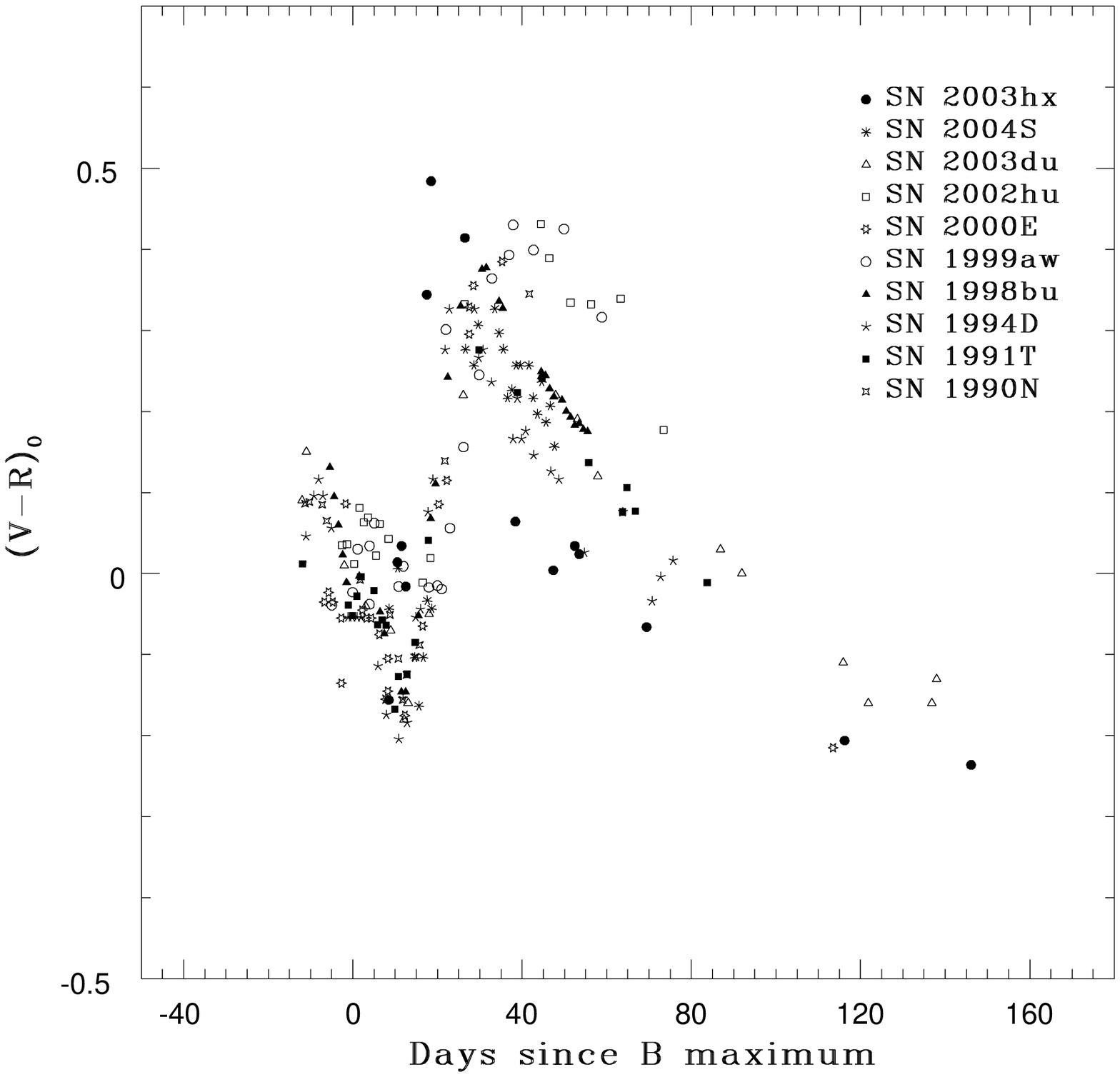}
\caption{The intrinsic $V-R$ color evolution of SN 2003hx compared with other supernovae of
type Ia.}
\label{vr_color}
\end{figure}

\begin{figure}
\centering
\includegraphics[height=8.0cm,width=8.0cm]{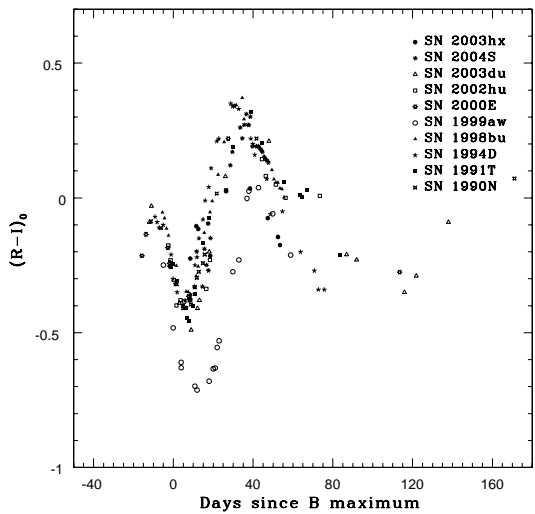}
\caption{The intrinsic $R-I$ color evolution of SN 2003hx compared with other supernovae of
type Ia.}
\label{ri_color}
\end{figure}

Supernovae of type Ia show significant uniformity in their intrinsic colors in late epochs
after maximum light. The dereddened $(B-V)$, $(V-R)$ and $(R-I)$ color curves of SN 2003hx are
shown in Figures \ref{bv_color}, \ref{vr_color} and \ref{ri_color}. The color curves
of SN 2003hx are dereddened using the total extinction values listed in Table \ref{parameters}
and discussed in section \ref{reddening}. For a comparison, we show here the color
curves of other type Ia supernovae which were reddening corrected using Cardelli extinction
law (Cardelli, Clayton \& Mathis 1989) and the $E(B-V)$ values of
$E(B-V)$ = 0.026 (galactic reddening only using Schlegel, Finkbeiner \& Davis 1998) for SN 1990N,
$E(B-V)$ = 0.13 for SN 1991T (Phillips et al. 1992),
$E(B-V)$ = 0.04 for SN 1994D (Richmond et al. 1995),
$E(B-V)$ = 0.32 for SN 1998bu (Hernandez et al. 2000),
$E(B-V)$ = 0.032 for SN 1999aw (Strolger et al. 2002),
$E(B-V)$ = 0.5 for SN 2000E (Valentini et al. 2003),
$E(B-V)$ = 0.044 for SN 2002hu (Sahu et al. 2006),
$E(B-V)$ = 0.02 for SN 2003du (Anupama et al. 2005) and
$E(B-V)$ = 0.18 for SN 2004S (Misra et al. 2005).

The evolution of $(B-V)_0$ color of SN 2003hx is significantly different than other type Ia
supernovae, however the overall trend 40 days past $B$ maximum is the same as other type Ia
supernovae. At the epoch of $B$ band maximum, $(B-V)_0$ = 0.02 mag which is consistent
with the intrinsic colors at maximum observed for other type Ia supernovae usually in the
range of $\sim$ -0.1 to +0.1 mag. Type Ia supernovae attain $(B-V)_0$ = 1 mag $\sim$ 30 days
after $B$ maximum. We do not have color estimate at $\sim$ 30 days post $B$ band maximum, however
$(B-V)_0$ = 0.85 at $\sim$ 38 days after $B$ maximum. After this epoch the color gets bluer as also
seen for other supernovae of type Ia. The overall $(B-V)_0$ color evolution of SN 2003hx
is not bluer as seen in Figure \ref{bv_color} for other supernovae.

The $(V-R)_0$ color curves of SN 2003hx also evolves in a manner similar to other SNe Ia. However,
the colors of SN 2003hx are quite similar to SN 1999aw till $\sim$ 20 days after $B$ band maximum.
The early $(V-R)_0$ color is bluest at -0.16 mag $\sim$ 8 days past $B$ band maximum although
bluer $(V-R)_0$ color is observed after 80 days after $B$ maximum. $(V-R)_0$ reaches 0.48 mag after
$\sim$ 18 days after $B$ maximum which is quite different as compared to other typical type Ia
supernovae (Figure \ref{vr_color}).

The $(R-I)_0$ color evolution of SN 2003hx follows the similar trend as seen for other type Ia
supernovae compared here. The evolution of $(R-I)_0$ color of SN 2003hx till $\sim$ 30 days after
$B$ maximum is similar to other type Ia supernovae, but after 30 days the $(R-I)_0$ colors are
generally bluer than the other supernovae. The $(R-I)_0$ color reaches its bluest value of -0.23 mag
around 8 days after $B$ maximum and subsequently gets redder after that.

Comparing $(B-V)_0$, $(V-R)_0$ and $(R-I)_0$ colors of different type Ia supernovae, we see that the
$(B-V)_0$ and $(R-I)_0$ color evolution of SN 2003hx is quite close to the respective colors of SN 1999aw.

\begin{figure}
\centering
\includegraphics[height=8.0cm,width=8.0cm]{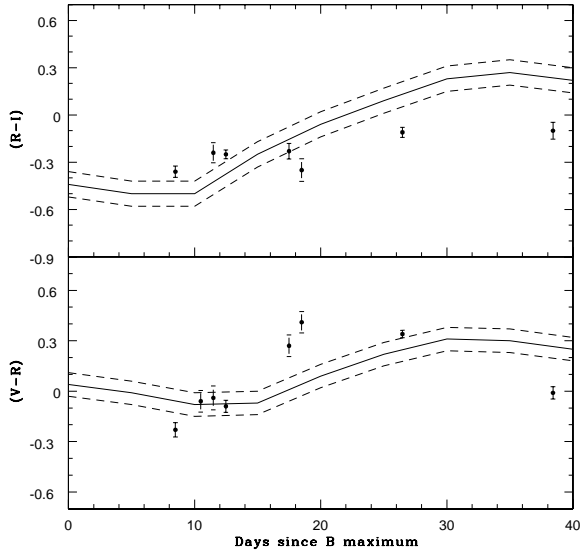}
\caption{(V-R) and (R-I) Color curves of SN 2003hx. Filled circles represent colors obtained from
our observations which have been corrected for the best fit value of selective extinction. The middle
(solid) line in each panel shows the intrinsic color curves of Nobili et al. (2003) bounded by errors
on both sides (dashed line).}
\label{nobili_colors}
\end{figure}

We also compare the colors of SN 2003hx with those obtained using the intrinsic color curves of
the SNe Ia population given by Nobili et al. (2003). Nobili et al. (2003) present the intrinsic color
curves of 48 type Ia SNe till 40 days after $B$ band maximum. The total selective extinction
was estimated,taking it as a fit parameter,by comparing the observed colors of SN 2003hx with
the intrinsic colors given by Nobili et al. (2003).
The overall shape of the observed $(V-R)$ and $(R-I)$ color curves are similar to the intrinsic color
curves (Figure \ref{nobili_colors}), except for systematic shifts in individual colors due to selective
extinction. Best fitting value of selective extinction obtained are $E(V-R) = 0.32 \pm 0.07$ and
$E(R-I) = 0.45 \pm 0.08$. Reddening estimates from other methods are discussed in
section \ref{reddening} and listed in Table \ref{parameters}.

\subsection{Comparison of the light curves}
\begin{figure}
\centering
\includegraphics[height=8.0cm,width=8.0cm]{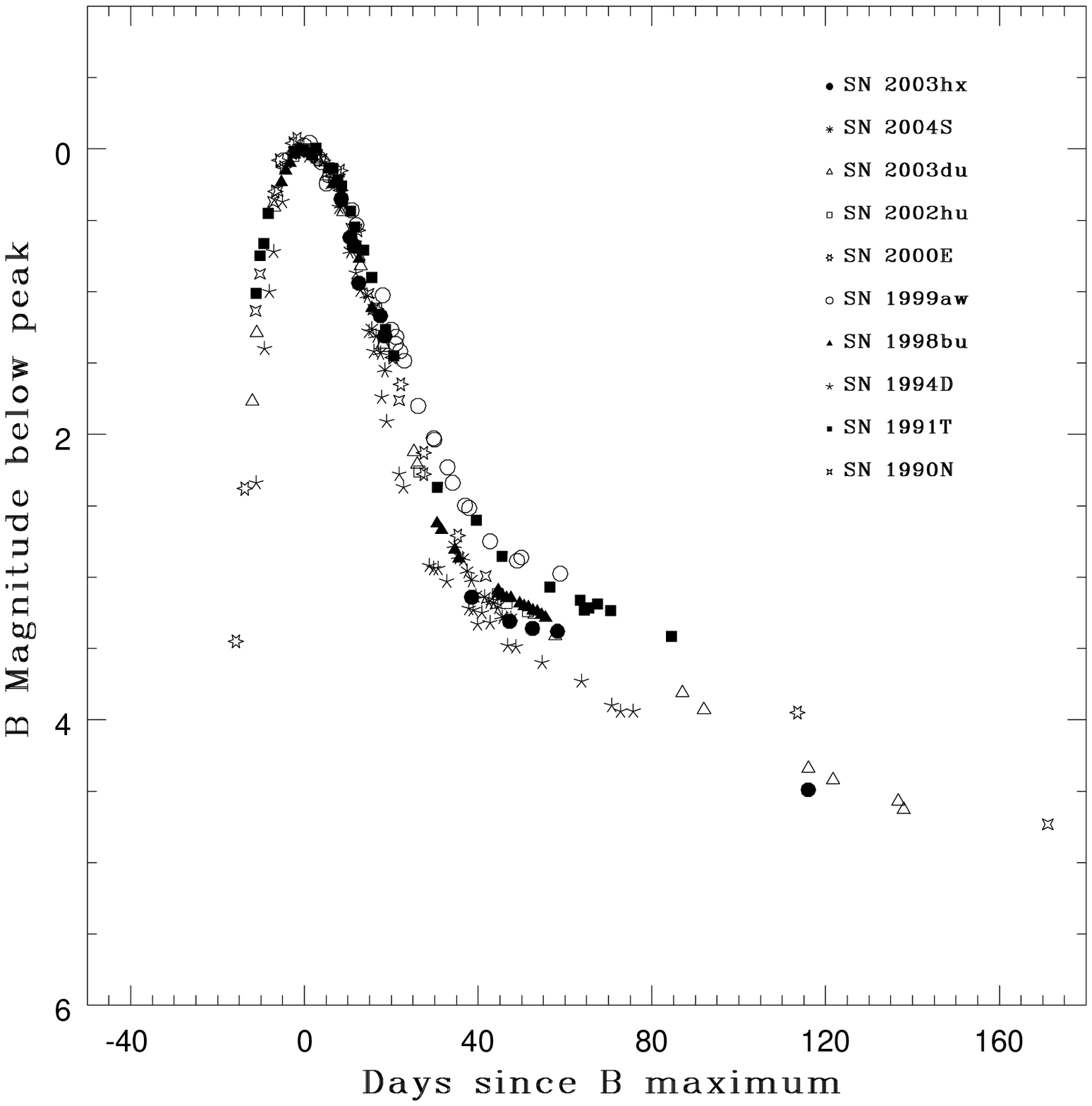}
\caption{B band light curve of SN 2003hx together with those of SN 2004S, SN 2003du, SN 2002hu,
SN 2000E, SN 1999aw, SN 1998bu, SN 1994D, SN 1991T and SN 1990N. All the light curves are shifted
to match the time of $B$ maximum and peak magnitude in $B$ band.}
\label{sn2003hx_blc_compare}
\end{figure}

\begin{figure}
\centering
\includegraphics[height=8.0cm,width=8.0cm]{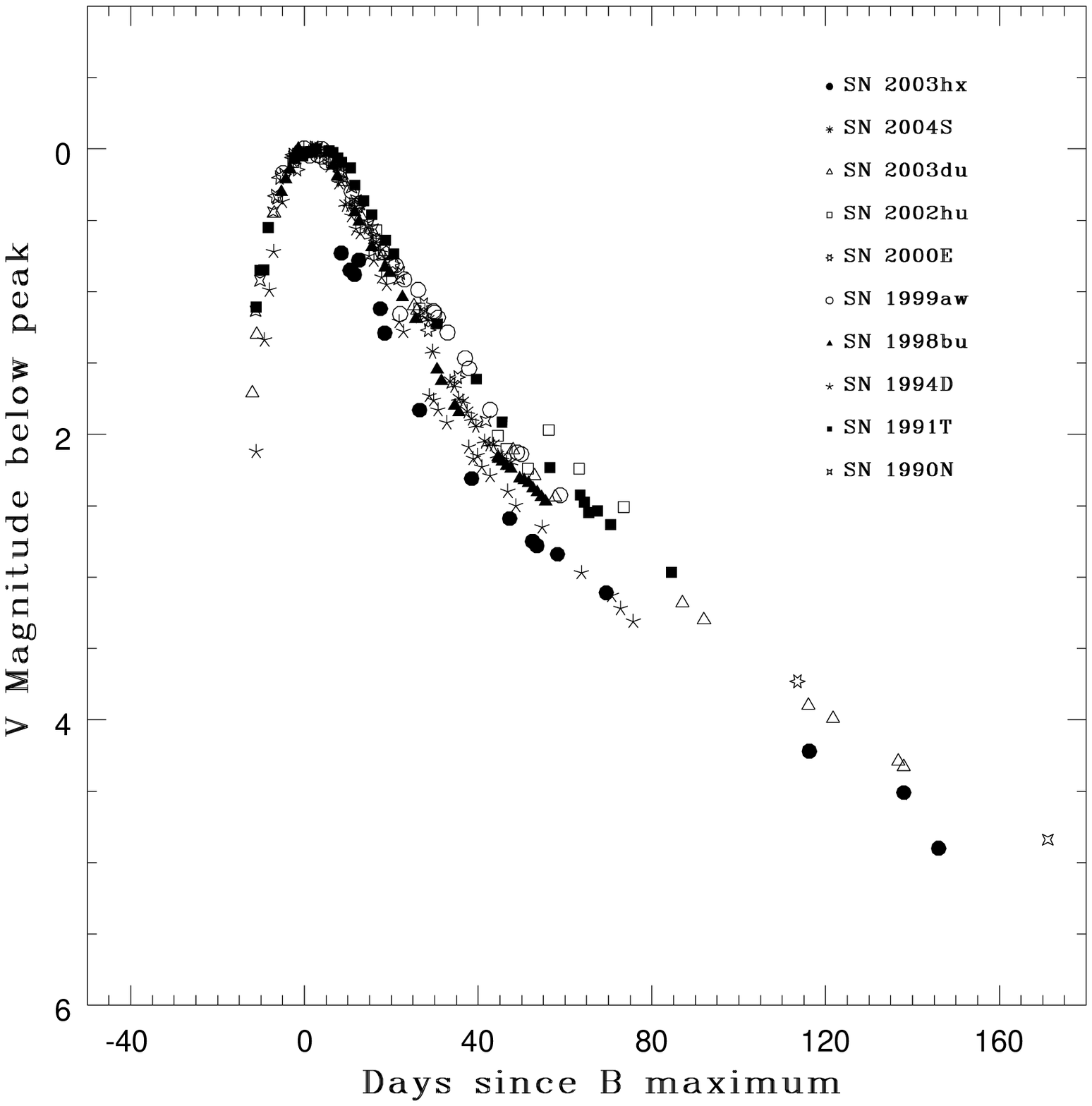}
\caption{$V$ band light curve of SN 2003hx together with those of SN 2004S, SN 2003du, SN 2002hu,
SN 2000E, SN 1999aw, SN 1998bu, SN 1994D, SN 1991T and SN 1990N. All the light curves are shifted
to match the time of $B$ maximum and peak magnitude in $V$ band.}
\label{sn2003hx_vlc_compare}
\end{figure}

\begin{figure}
\centering
\includegraphics[height=8.0cm,width=8.0cm]{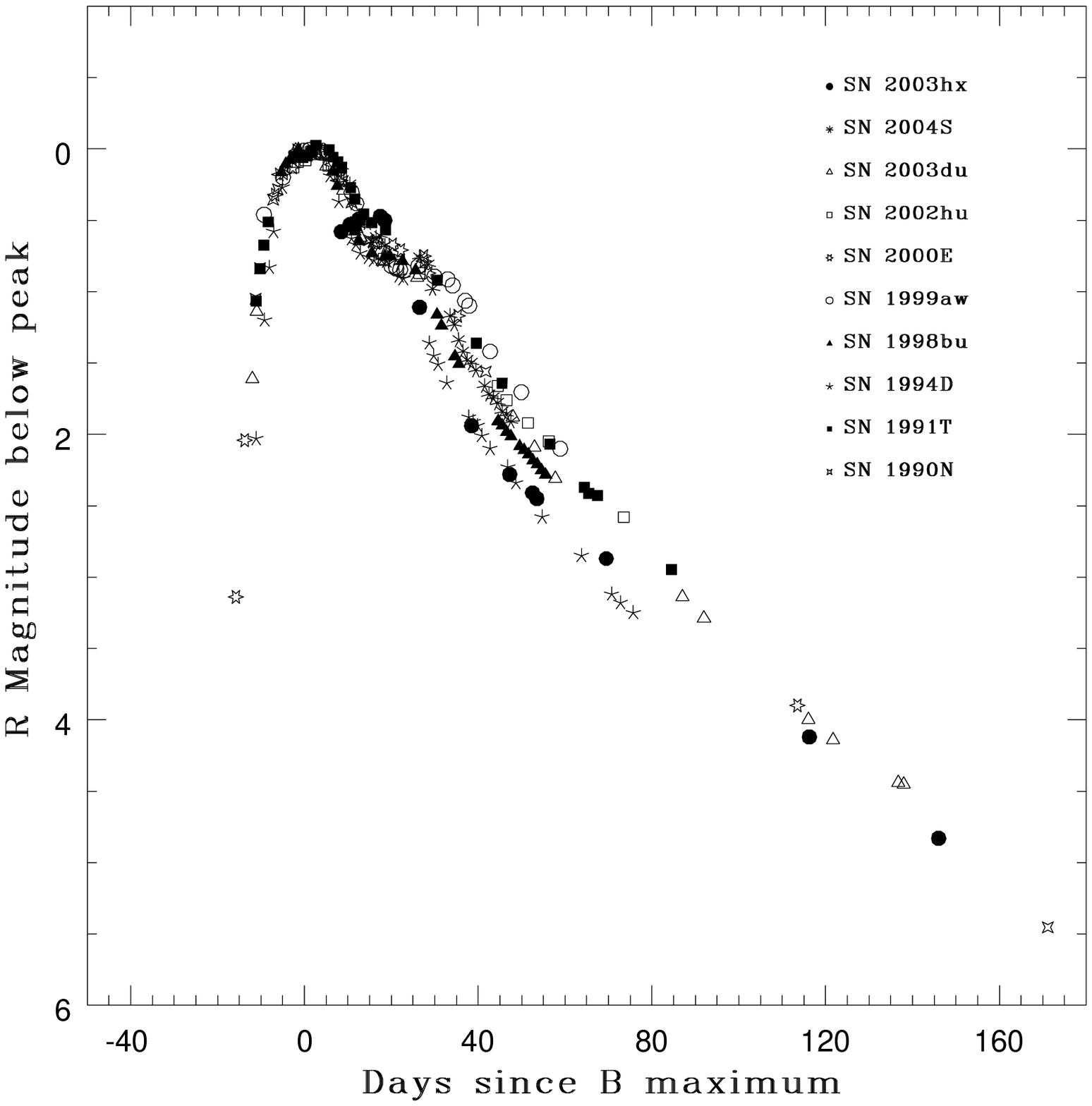}
\caption{$R$ band light curve of SN 2003hx together with those of SN 2004S, SN 2003du, SN 2002hu,
SN 2000E, SN 1999aw, SN 1998bu, SN 1994D, SN 1991T and SN 1990N. All the light curves are shifted
to match the time of $B$ maximum and peak magnitude in $R$ band.}
\label{sn2003hx_rlc_compare}
\end{figure}

\begin{figure}
\centering
\includegraphics[height=8.0cm,width=8.0cm]{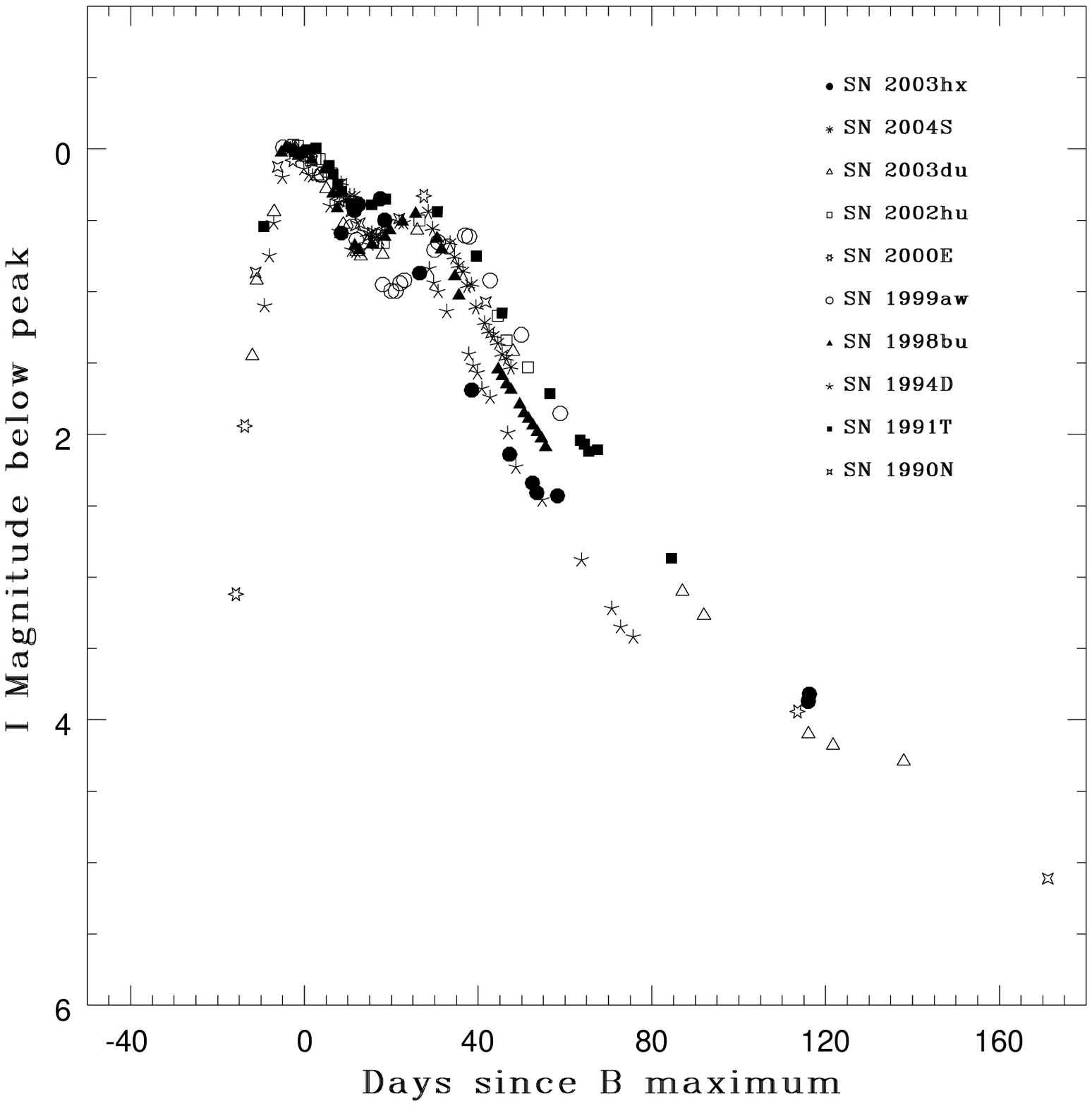}
\caption{$I$ band light curve of SN 2003hx together with those of SN 2004S, SN 2003du, SN 2002hu,
SN 2000E, SN 1999aw, SN 1998bu, SN 1994D, SN 1991T and SN 1990N. All the light curves are shifted
to match the time of $B$ maximum and peak magnitude in $I$ band.}
\label{sn2003hx_ilc_compare}
\end{figure}

In Figures \ref{sn2003hx_blc_compare}, \ref{sn2003hx_vlc_compare}, \ref{sn2003hx_rlc_compare} and
\ref{sn2003hx_ilc_compare} we compare the $BVRI$ light curves respectively of SN 2003hx with other
normal type Ia SNe:
SN 1990N ($\Delta m_{15}(B)$ = 1.03, Lira et al. 1998),
SN 1991T ($\Delta m_{15}(B)$ = 0.95, Lira et al. 1998),
SN 1994D ($\Delta m_{15}(B)$ = 1.26, Richmond et al. 1995),
SN 1998bu ($\Delta m_{15}(B)$ = 1.01, Suntzeff et al. 1999),
SN 1999aw ($\Delta m_{15}(B)$ = 0.81, Strolger et al. 2002)
SN 2000E ($\Delta m_{15}(B)$ = 0.94, Valentini et al. 2003),
SN 2002hu ($\Delta m_{15}(B)$ = 1.00, Sahu et al., 2006),
SN 2003du ($\Delta m_{15}(B)$ = 1.04, Anupama et al. 2005) and
SN 2004S ($\Delta m_{15}(B)$ = 1.26, Misra et al. 2005).
All the light curves have been plotted by normalizing to the epoch of $B$ band maximum for each
supernova and the respective peak magnitudes in different bands. Comparison of the $BVRI$ light
curves of SN 2003hx shows that they are quite similar to other type Ia SNe till $\sim$ 20 days
after $B$ band maximum whereas at a later epoch they are fainter than other SNe Ia. The $B$ band
light curve is similar to SN 2004S 20 days after $B$ band maximum. Post 20 days after $B$ maximum,
the $VRI$ light curves of SN 2003hx are very similar to those of SN  1994D which is a normal
type Ia SNe with a decline rate of $\Delta m_{15}(B)$ = 1.26 mag. 80 days after $B$ maximum
SN 2003hx compares well with the light curves of SN 2003du.
Both Sn 2003hx and SN 1994D were hosted by lenticular galaxies NGC 2076 and NGC 4256 respectively.
The light curve parameter $\Delta m_{15}(B)$ for
different type Ia SNe is listed in Table \ref{comparison_sn}.

\begin{table*}
\caption{Parameters of SN 2003hx}
\medskip
\begin{center}
\begin{tabular}{ll} \hline\hline
&\\
Discovery date & 2003 September 12.5 UT\\
Host galaxy & NGC 2076\\
Galaxy type & Morphological type S0-a\\
RA (2000) & $05^{\rm h} 46^{\rm m} 46^{\rm s}.97$\\
Dec. (2000) & $-16^{\circ} 47^\prime 00^{\prime\prime}.6$\\
Offset from the nucleus & 5.2 arcsec W and 2.6 arcsec S \\
Spectrum & Type Ia\\
Radial velocity from galaxy redshift (km sec$^{-1}$) & 2142 $\pm$ 5 (NED)\\
& 2137 $\pm$ 7 (LEDA)\\
Radial velocity corrected for LG infall on to Virgo (km sec$^{-1}$) & 1967 (LEDA) \\
Expansion velocity of the SN & 12000 km sec$^{-1}$\\
Distance modulus & 32.456 mag\\
(H$_0$ = 65 km sec$^{-1} Mpc^{-1}$) & \\
Epoch of maximum (from template fitting) & $B$ band 245 2893.0 $\pm$ 1.0 \\
& $V$ band 245 2888.3 $\pm$ 0.5 \\
& $R$ band 245 2886.7 $\pm$ 1.0 \\
& $I$ band 245 2889.6 $\pm$ 0.6\\
Magnitude at maximum & $B$ = 14.92 $\pm$ 0.01 \\
& $V$ = 14.34 $\pm$ 0.03\\
& $R$ = 14.40 $\pm$ 0.06\\
& $I$ = 14.30 $\pm$ 0.03\\
Colors at maximum & $B-V$ = 0.58 $\pm$ 0.03\\
& $V-R$ = -0.06 $\pm$ 0.06\\
& $R-I$ = 0.10 $\pm$ 0.06\\
Absolute magnitudes at maximum & $M^B$ = -19.20 $\pm$ 0.18 \\
& $M^V$ = -19.22 $\pm$ 0.15\\
& $M^R$ = -18.91 $\pm$ 0.22\\
& $M^I$ = -18.70 $\pm$ 0.17\\
Reddening estimate: Using $\Delta m_{15}$ and intrinsic luminosity Phillip's relation & E(B-V) = 0.56 $\pm$ 0.23\\
Phillip's relation & E(B-V) = 0.64 $\pm$ 0.12\\
& E(V-I) = 0.35 $\pm$ 0.12\\
Lira's relation & E(B-V) = 0.43 $\pm$ 0.04\\
Using intrinsic colors of Nobili et al. (2003) at maximum & E(B-V) = 0.69 $\pm$ 0.03\\
& E(V-I) = 0.47 $\pm$ 0.04\\
$R_v$ = 1.97 $\pm$ 0.54\\
Adopted total extinction (mag) & $A_B$ = 1.664\\
& $A_V$ = 1.104\\
& $A_R$ = 0.858\\
& $A_I$ = 0.543\\
Magnitude of secondary $I$ maximum (from template fitting) & 14.62\\
$\Delta m_{15}$ in B from template & 1.11\\
$\Delta m_{15}$ in B from observations & 1.17 $\pm$ 0.12\\
Decline rate per day & $B$ band 0.092 $\pm$ 0.06\\
& $V$ band 0.051 $\pm$ 0.05\\
& $R$ band 0.072 $\pm$ 0.04\\
& $I$ band 0.059 $\pm$ 0.07\\
&\\
\hline
\end{tabular}
\label{parameters}
\end{center}
\end{table*}

%
\begin{table*}
\caption{Absolute $B$ magnitude at maximum, peak luminosity, $\Delta m_{15}(B)$ and $^{56}Ni$ masses
of type Ia supernovae.}
\medskip
\begin{center}
\begin{tabular}{lllllll} \hline \hline
SN & Galaxy & $M_B$ & $\Delta m_{15}(B)$ & log $L_{bol}$ & $M_{Ni}$ & Reference\\
   &        & (mag) &  (mag)             & (erg s$^{-1}$)& M$_\odot$& \\
\hline
&&&&&& \\
1990N  & NGC 4639      &                   & 1.03            &                 &                 & 1\\
1991T  & NGC 4527      & -20.06            & 0.95 $\pm$ 0.05 & 43.36           & 1.14            & 2,3\\
1994D  & NGC 4526      & -18.95 $\pm$ 0.18 & 1.31 $\pm$ 0.08 & 42.91           & 0.41            & 2,4\\
1998bu & NGC 3368      & -19.67 $\pm$ 0.20 & 1.01 $\pm$ 0.05 & 43.18           & 0.77            & 2,5\\
1999aw & Anonymous     & -19.48 $\pm$ 0.11 & 0.81 $\pm$ 0.03 & 43.18           & 0.76            & 6\\
2000E  & NGC 6951      & -19.45            & 0.94 $\pm$ 0.05 & 43.29           & 0.90 $\pm$ 0.20 & 7\\
2002hu & MCG+6-6-12    & -19.38 $\pm$ 0.30 & 1.00 $\pm$ 0.05 & 43.25 $\pm$ 0.07&                 & 8 \\
2003du & UGC 9391      & -19.34 $\pm$ 0.30 & 1.04 $\pm$ 0.04 & 43.14           & 0.88            & 9\\
2004S  & MCG-05-16-021 & -19.05 $\pm$ 0.23 & 1.26 $\pm$ 0.06 & 42.94           & 0.41            & 10\\
{\bf 2003hx} & {\bf NGC 2076}   & {\bf -19.20 $\pm$ 0.18} & {\bf 1.17 $\pm$ 0.12} & {\bf 43.01} & {\bf 0.66}& 11\\
&&&&&& \\
\hline
\end{tabular}
\label{comparison_sn}
\end{center}
{\it References}: 1-Lira et al. 1998, 2-Contardo, Leibundgut \& Vacca 2000, 3-Lira et al. 1998, 4-Richmond et al. 1995,
5-Suntzeff et al. 1999, 6-Strolger et al. 2002, 7-Valentini et al. 2003, 8-Sahu et al. 2006, 9-Anupama et al. 2005,
10-Misra et al. 2006, 11-Present work
\end{table*}

\section{Reddening estimate}
\label{reddening}
\begin{figure*}
\centering
\includegraphics[height=14.0cm,width=14.0cm]{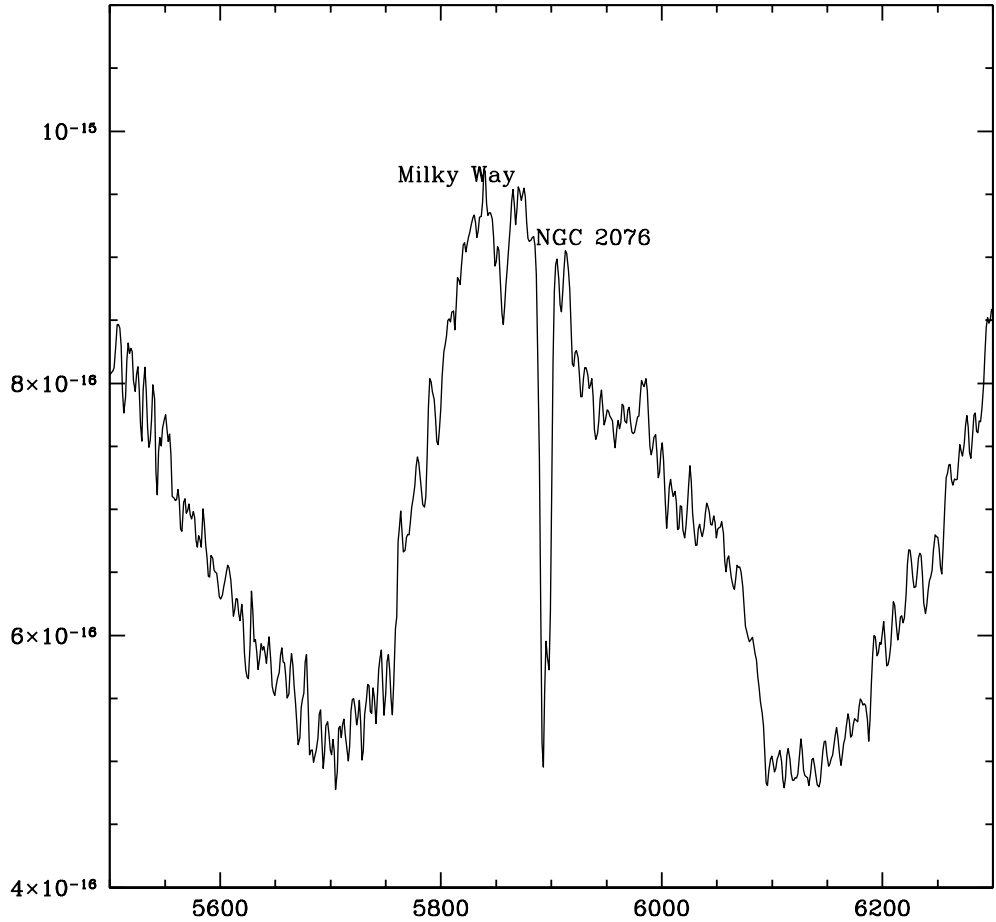}
\caption{Spectra of SN 2003hx around  NaID absorption line. The two components due to the Milky Way
and the host galaxy is clearly seen.}
\label{ext}
\end{figure*}

The estimated reddening in the direction of SN 2003hx due to our own galaxy from Schlegel,
Finkbeiner \& Davis (1998) is E(B-V) = 0.084 mag. The supernova occurred very close to the
nucleus of the galaxy NGC 2076, we therefore expect substantial reddening due to the host
galaxy. The optical spectra demonstrates that SN 2003hx is a highly reddened supernova (refer Section \ref{spectra}).
Figure \ref{ext} shows the strong interstellar NaID lines at the rest wavelength of the host galaxy
and relatively weak absorption line due to the extinction in the Milky Way.
The average equivalent width of the Galactic component of the NaID absorption is 0.54$\pm$0.11 \AA,
whereas the equivalent width of NaI D absorption due to the host galaxy is 5.02$\pm$0.26 \AA.
Using the two relations between equivalent width of NaID line and $E(B-V)$
(Turatto, Benetti \& Cappellaro 2003) the measured equivalent width implies
$E(B-V)$ as 0.09 and 0.28 for the Milky Way, the lower value being close to the estimate of
Schlegel, Finkbeiner \& Davis (1998). The $E(B-V)$ values due to the host galaxy
is estimated as  0.80 and 2.56. The sum of the lower values of $E(B-V)$ for Milky Way and
host galaxy gives total reddening as  0.89.
Further,  we calculate $E(B-V)$ at the time of maximum using intrinsic color at the
epoch of maximum given by Nobili et al. (2003) which is listed in Table \ref{parameters}.
We also estimate the total extinction following the photometric methods of Phillips et al.
(1999) and the Lira's method (1995).
However, the best estimate of reddening comes from a good measurement of $\Delta m_{15}$
from where we deduce the intrinsic luminosity. This relationship between $\Delta m_{15}$
and intrinsic luminosity is much more exhaustively tested than the extinction law. We obtain,
using this method, our independent measure of $R_v$ towards SN 2003hx. The value of $R_v$ is listed
is listed in Table \ref{parameters}.
The values of total selective extinction obtained
from different methods is listed in Table \ref{parameters}.
Wang \& Baade (2003)
based on the spectropolarimetric observations inferred an equivalent width of 4.86 \AA for the
interstellar Na I D line, indicating significant dust extinction and a polarization of
$\sim$ 2 percent. If this polarization is due to dust in the host galaxy, it implies that
the dust particles are smaller in size than their galactic counterparts. A similar conclusion
about the dust particle size was  arrived at by Sahu et al. (1998), in their study of dust
property of the host galaxy NGC 2076.
Our independent estimates of $R_v$ also indicates the small size of dust particles as
compared to their galactic counterparts.
Wand \& Baade (2003)
based on these observations, found the ratio of total to selective extinction $R_v = A_v/E(B-V)$
to be 2.2. We adopt $E(B-V)$ = 0.56 $\pm$ 0.23 (using the value obtained from the relation
between $\Delta m_{15}$ and intrinsic luminosity) and $R_v$ = 1.97 $\pm$ 0.54 to estimate
the total extinction in different filters, the values of which are listed in
Table \ref{parameters}.
We see that the reddening due to our own galaxy is very small
in comparison to the total reddening. Thus, a large amount of extinction could arise in the
host galaxy of SN 2003hx.

\section{Absolute luminosity and Bolometric Light Curve}
\label{bolometric}
Assuming $H_0$ = 70 km sec$^{-1}$ Mpc${-1}$ and the radial velocity of NGC 2076 as
$v_r$ = 2142 $\pm$ 5 km sec$^{-1}$, we find a distance modulus of 32.456 mag. The
total extinction estimated is mentioned in Table \ref{parameters}. From these the
absolute magnitudes estimated in different bands are M$^B$ = -19.20 $\pm$ 0.18,
$M^V$ = -19.22 $\pm$ 0.15, $M^R$ = -18.91 $\pm$ 0.22 and $M^I$ = -18.70 $\pm$ 0.17.
Altavilla et al. (2004) suggest another method for estimating absolute magnitude
using a relation between M$_{max}$ and $\Delta m_{15}$. Adopting the values of
linear fit coefficients as given by Altavilla et al. (2004) for $R_v$ = 2.2, we
obtain $M^B_{max}$ = -19.38 $\pm$ 0.10. The values of the absolute magnitude in $B$ band
obtained by the above two methods are in good agreement with each other.

\begin{figure}
\centering
\includegraphics[height=8.0cm,width=8.0cm]{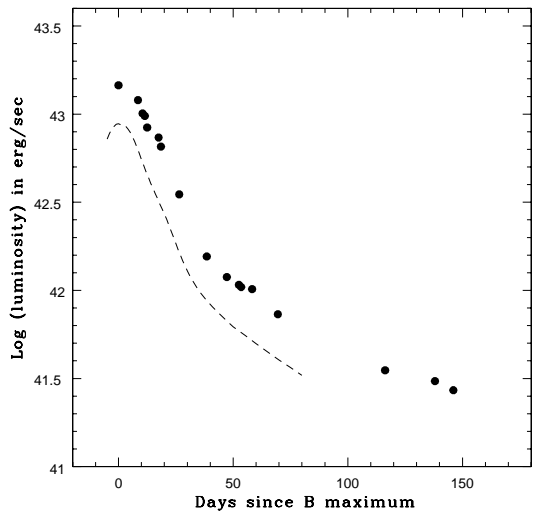}
\caption{Bolometric light curve of SN 2003hx. The dots show the corrected $UVOIR$ bolometric
light curve and the dash line shows the bolometric light curve constructed from the $BVI$ fits
of Hamuy et al. (1996).}
\label{bolometric_lc}
\end{figure}

The bolometric light curve of SN 2003hx is estimated using the optical observations
presented here. During the early phase most of the flux emerges in the optical from a
type Ia supernova (Suntzeff 1996), thus the integrated flux in $UBVRI$ bands gives a
good estimate of the bolometric luminosity. The peak bolometric luminosity is directly
related to the radioactive Nickel ejected in the explosion. The dereddened magnitudes
were converted to flux using calibrations by Fukugita, Shimasaku \& Ichikawa (1995).
Since we have very few $U$ band observations, we corrected for the missing passband flux
in the optical for a contribution of 10 percent as shown by Contardo et al. (2000). Also,
to construct the full $UVOIR$ bolometric light curve we should combine the ultraviolet
and the near-IR data with the optical data. Suntzeff (1996) constructed the full $UVOIR$
bolometric light curve for SN 1992A by integrating the flux in wavelength range of 2000 $\AA$
to 2.2 $\mu$m. This shows that both the ultraviolet and the near-IR contribution is
$\sim$ 10 percent each of the total $UVOIR$ luminosity for $\sim$ 80 days since the time
of $B$ band maximum. Thus, a correction for the missing flux is required and has to be
applied to the bolometric luminosity which is estimated using the optical data alone. We
correct for a total contribution of 20 percent from both ultraviolet and near-IR regions.
In Figure \ref{bolometric_lc} we show the $UVOIR$ bolometric light curve as dots. The dash
line in Figure \ref{bolometric_lc} shows the contribution derived from the $BVI$ bands alone, as obtained
from fitted templates from -5 to 80 days with reference to the time of $B$ band maximum.
The $UVOIR$ luminosity estimated indicates a peak luminosity of logL = 43.01.
The light curves are powered by radioactivity and the amount of $^{56}$Ni mass ejected may be estimated
using the peak luminosity (Arnett 1982).
Assuming a rise-time of $\sim$ 18 days for SN 2003hx and the peak $UBVRI$ bolometric luminosity determined,
the amount of $^{56}$Ni is estimated to be $M_{Ni}$ = 0.50 $M_\odot$. Using the corrected bolometric
luminosity, the $^{56}$Ni mass estimate is 0.66 $M_\odot$.

\begin{figure*}
\centering
\includegraphics[height=13.0cm,width=10.0cm,angle=-90]{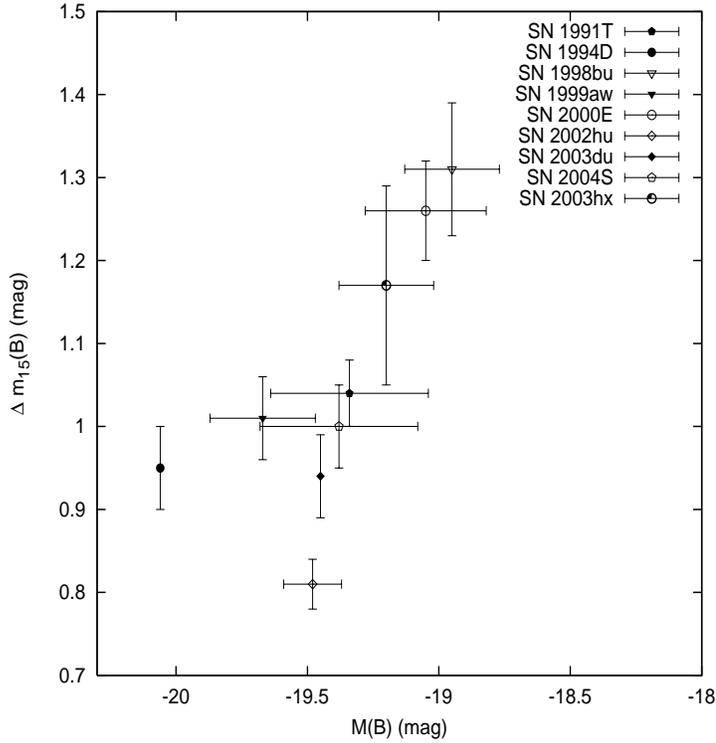}
\caption{Comparison of maximum absolute B band magnitudes and $\Delta m_{15}$ of SN 2003hx with other type Ia supernovae.}
\label{m_bvsdel_m15}
\end{figure*}

\begin{figure*}
\centering
\includegraphics[height=13.0cm,width=10.0cm,angle=-90]{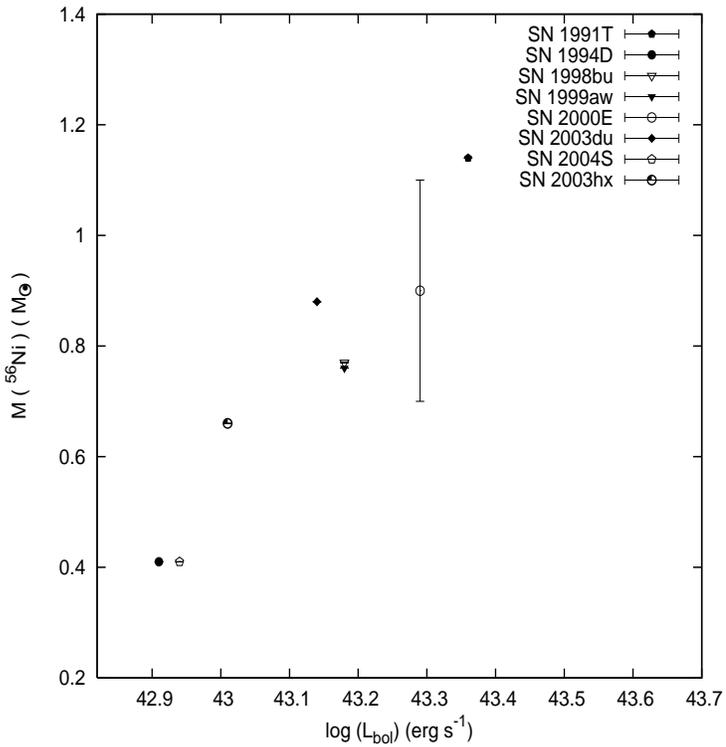}
\caption{Comparison of peak bolometric luminosity and ejected $^{56}Ni$ mass of SN 2003hx with other type Ia supernovae.}
\label{lbolvsm_ni}
\end{figure*}

We see that the peak bolometric luminosity and the ejected $^{56}$Ni mass of SN 2003hx is comparable
to that of SN 1998bu and SN 1999aw though the amount of $^{56}$Ni ejected in the other two cases is
slightly higher than SN 2003hx.
Table \ref{comparison_sn} lists the parameters of different type Ia supernovae including SN 2003hx. We
present plots of two simple parameters: absolute magnitude versus $\Delta m_{15}$ (Figure \ref{m_bvsdel_m15})
and log(L$_{bol}$) versus $^{56}$Ni (Figure \ref{lbolvsm_ni}).
We see that M$_{B}$ and $\Delta m_{15}$ as well as log(L$_{bol}$) and
$^{56}$Ni for type Ia supernovae follow a linear relation and the location of SN 2003hx in both these
plots agrees well with the rest of the sample.

\section{Spectral Evolution}
\label{spectra}
\begin{figure*}
\centering
\includegraphics[height=14.0cm,width=14.0cm]{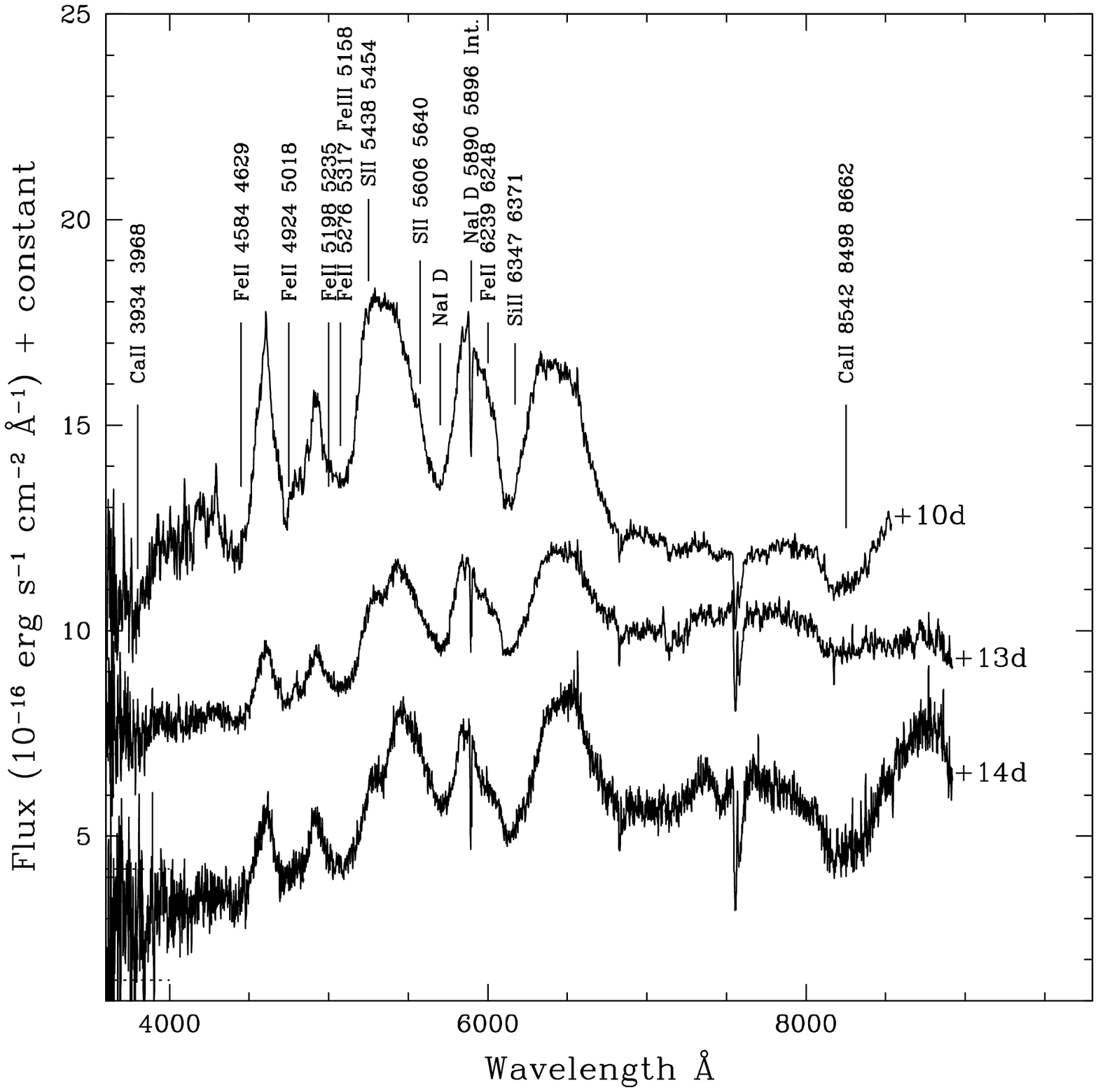}
\caption{The spectrum of SN 2003hx on +10, +13 and +14 days past the
$B$ maximum.}
\label{spec1}
\end{figure*}
\begin{figure*}
\centering
\includegraphics[height=14.0cm,width=14.0cm]{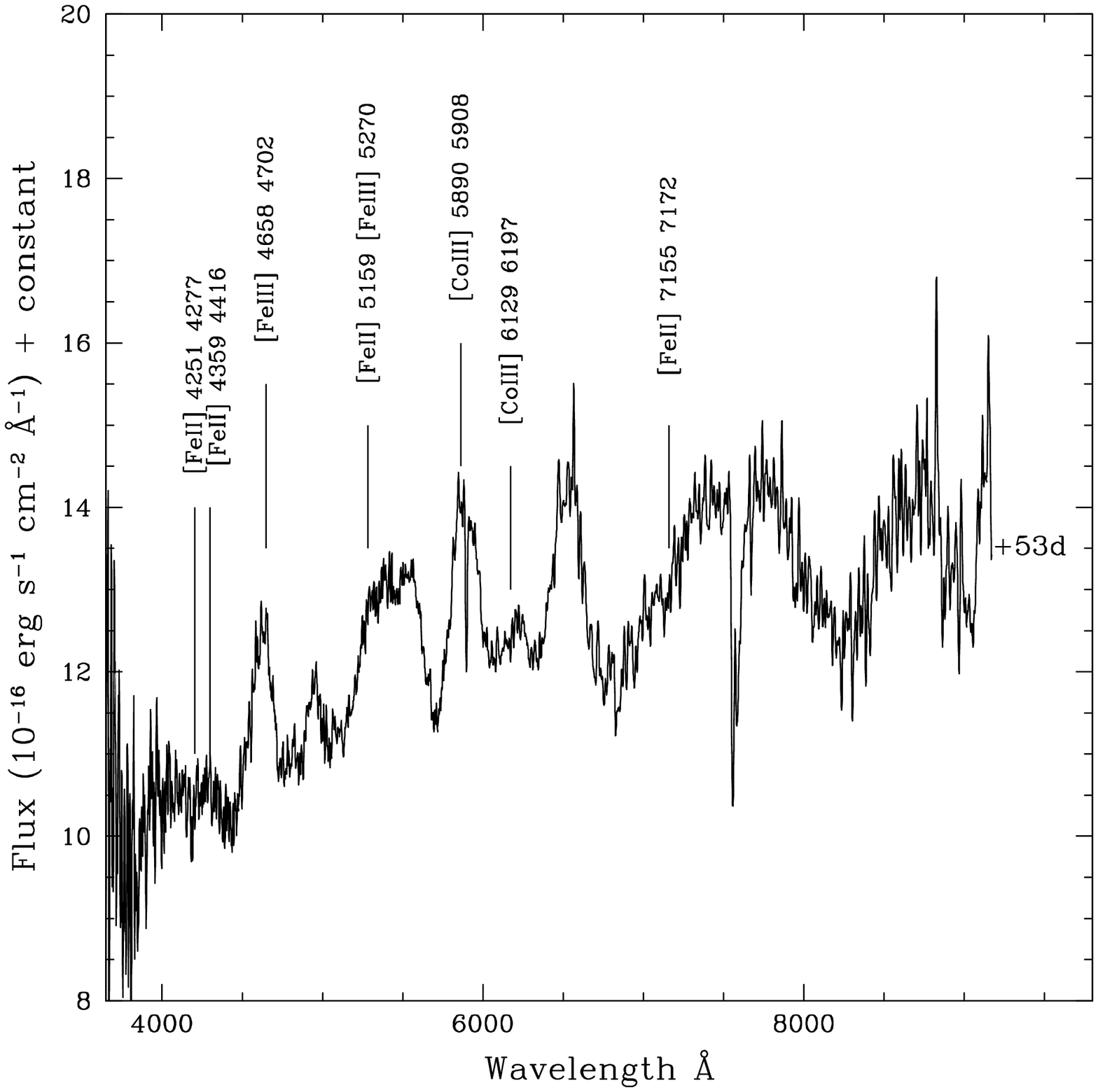}
\caption{The spectrum of SN 2003hx +53 days past $B$ maximum.}
\label{spec2}
\end{figure*}

 Optical spectra of SN 2003hx were obtained on four different epochs during
+10 days -- +53 days past the $B$ maximum. Figures \ref{spec1} and \ref{spec2}
show the spectral evolution of SN 2003hx. All the spectra show a deep
absorption around 6100\AA \ due to SiII, indicating the supernova to be a
normal type Ia event. The NaI D lines are clearly visible and strong,
suggesting a high value of reddening, consistent with the estimates in Section
\ref{reddening}.

\begin{figure*}
\centering
\includegraphics[height=14.0cm,width=14.0cm]{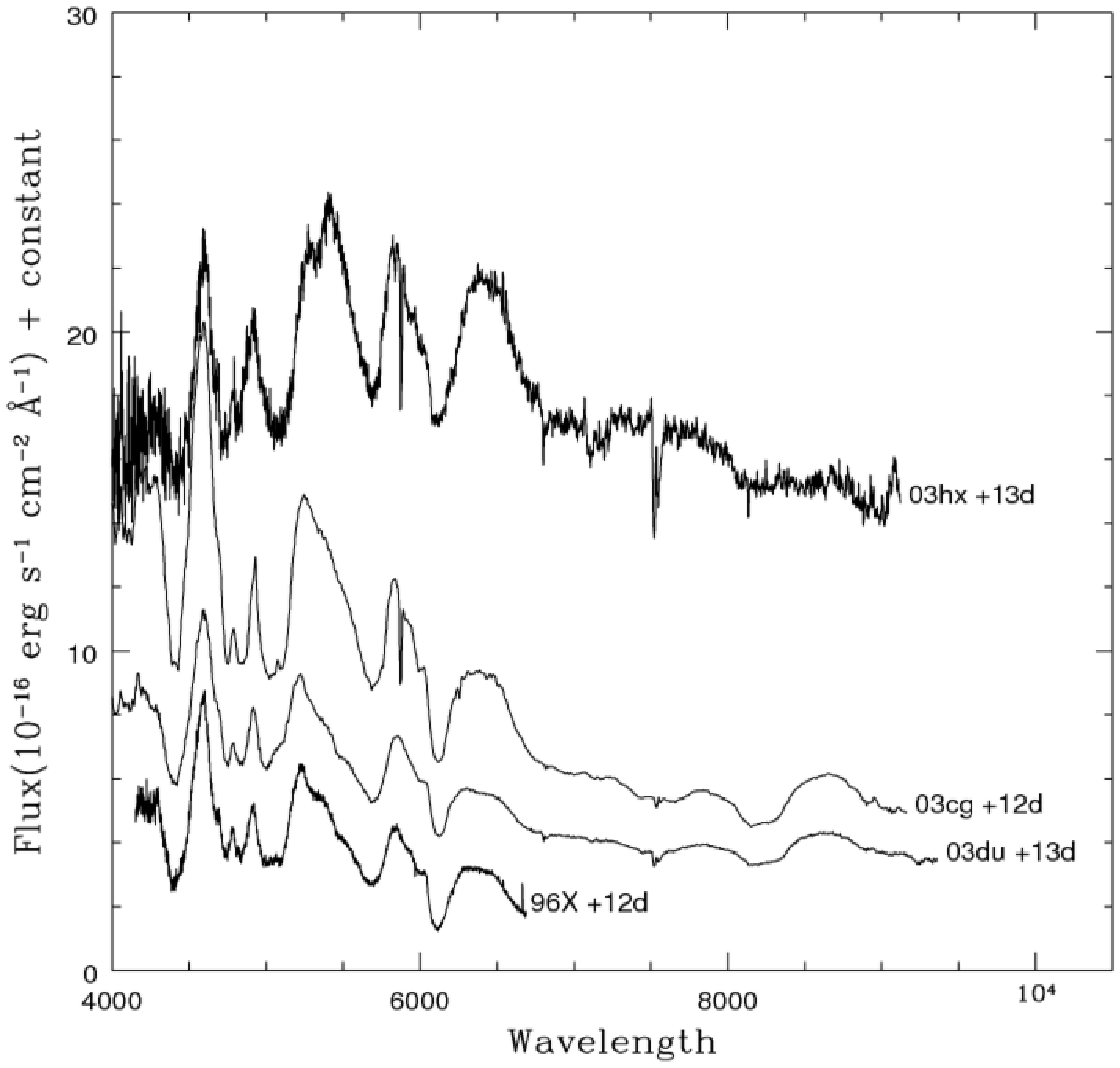}
\caption{Comparison of the spectrum of SN 2003hx on day +13 with those of
SN 2003du, SN 2003cg and SN 1996X at a similar epoch (see text for references).}
\label{comp_early}
\end{figure*}

\begin{figure*}
\centering
\includegraphics[height=14.0cm,width=14.0cm]{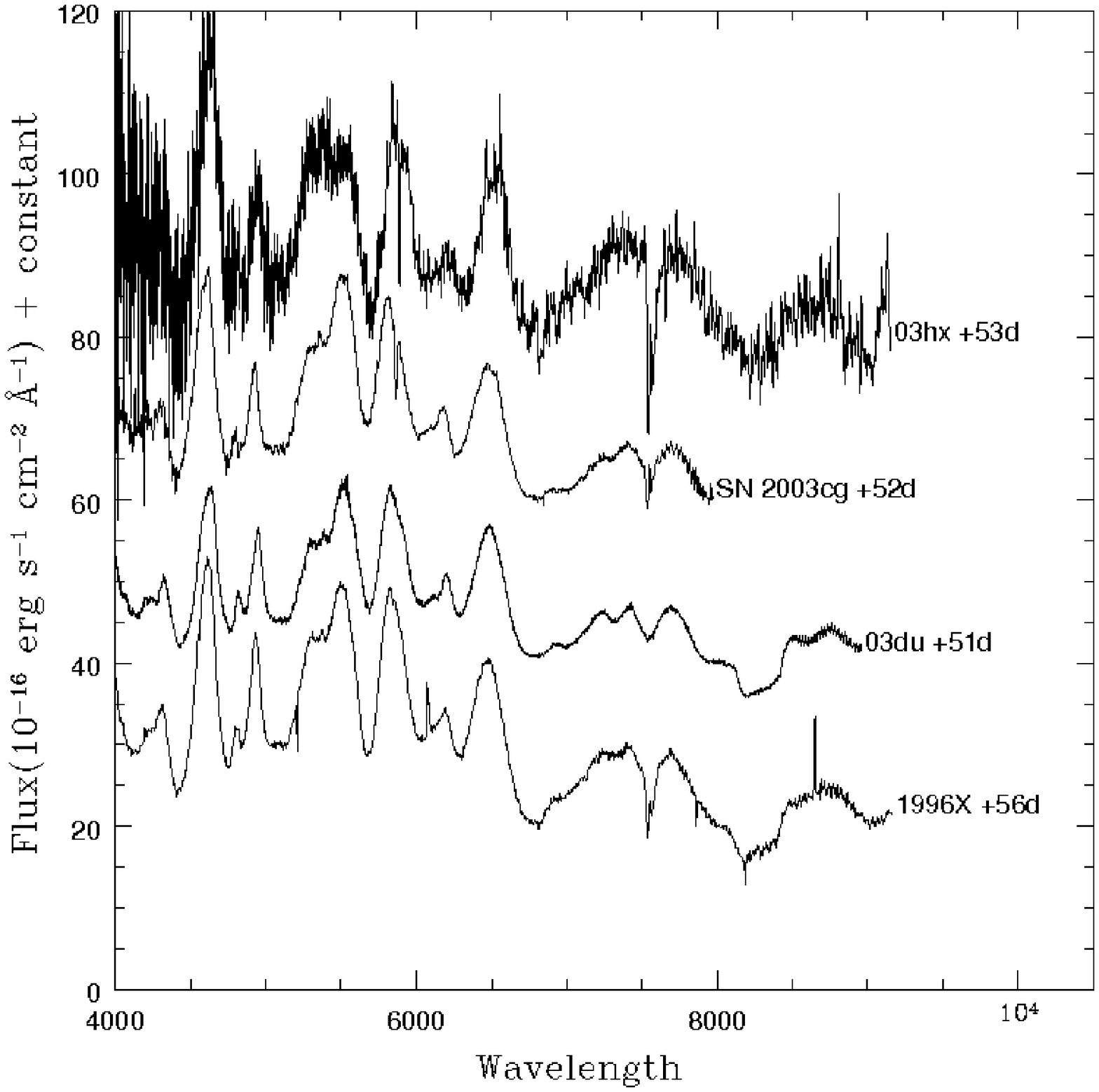}
\caption{Comparison of the spectrum of SN 2003hx on day +53 with those of
SN 2003du, SN 2003cg and SN 1996X at a similar epoch (see text for references).}
\label{comp_late}
\end{figure*}
A comparison of the spectra of SN 2003hx with those of other normal type Ia
events, namely, SN 2003cg (Elias-Rosa et al. 2006), SN 2003du (Anupama et al.
2005), and SN 1996X (Salvo et al. 2001) indicates an overall similarity
in the spectra. However, there are a few discrepancies. The +13d spectrum
(Figure \ref{comp_early} indicates very weak or no Ca II IR triplet in
SN 2003hx, while it is present, but weak compared to the other SNe at later
phases (Figure \ref{comp_late}). Further, the S II features at 5300--5600 \AA\
appear to be stronger in SN 2003hx.

\begin{figure*}
\centering
\includegraphics[height=14.0cm,width=14.0cm]{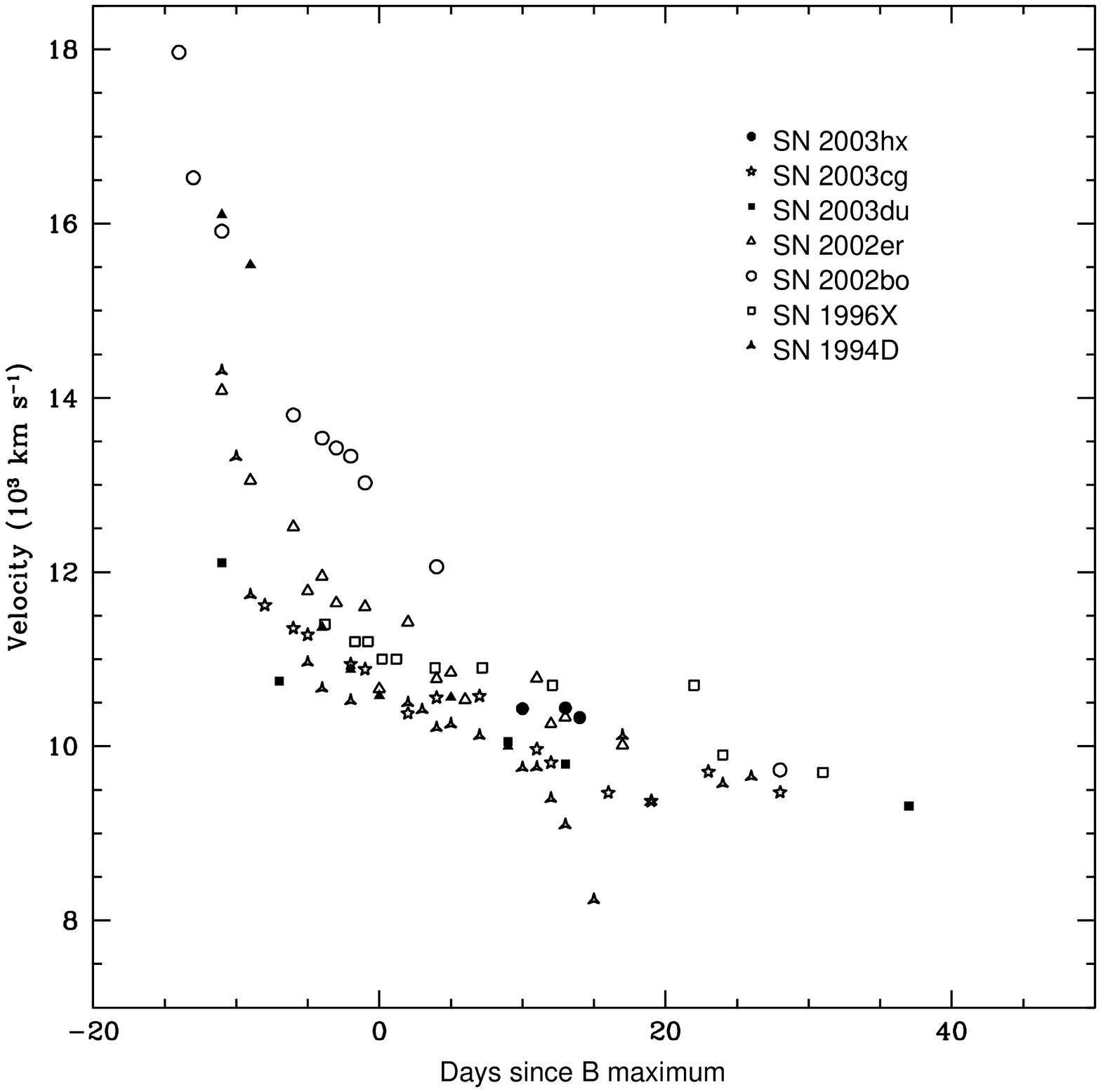}
\caption{Expansion velocity derived from the absorption minima of SiII 6355\AA\
line for SN 2003hx, SN 2003cg, SN 2003du, SN 2002er, SN 2002bo, SN 1996X
and SN 1994D (see text for references).}
\label{vel}
\end{figure*}

The expansion velocity is measured for the first three epochs, based on the
absorption minimum of Si II 6355 \AA\ line, and is found to be $\sim 10,400$
km sec$^{-1}$. As our spectroscopic coverage is rather sparse, not much can
be inferred about the velocity evolution in SN 2003hx. However, as seen in
Figure \ref{vel}, the velocity estimated for the first three epochs compares
well with the estimates for other type Ia events, SN 2003cg (Elias-Rosa et al.
2006), SN 2003du (Anupama et al. 2005), SN 2002er (Kotak et al. 2005),
SN 2002bo (Benetti et al. 2004), SN 1996X (Salvo et al. 2001) and SN 1994D
(Patat et al. 1996).

\section{Conclusions}
\label{conclusion}
We present here the $UBVRI$ photometric observations of SN 2003hx over a period of $\sim$ 146 days
after $B$ maximum obtained using the 2.01-m Himalayan Chandra Telescope (HCT). We also present
the optical spectra of SN 2003hx at four epochs. We study the light curve evolution in
$BVRI$ bands and estimate the peak magnitudes and the time of maximum in different bands
using the template fitting method. The light curve of SN 2003hx in the $B$ band is
approximated well with the template of SN 1992al whereas the $V$ and $I$ bands fit best with
the template of SN 1992A. The light curve parameter $\Delta m_{15}(B)$ is estimated to be
1.17 $\pm$ 0.12 which shows SN 2003hx as a mid-decliner. The colors of SN 2003hx are
quite close to the color estimates of SN 1999aw whereas the light curve evolution matches
very well with that of SN 1994D. Both SN 2003hx and SN 1994D occurred in lenticular
galaxies.
We infer from both photometric and spectroscopic studies that SN 2003hx is a highly reddened supernova
with $E(B-V)$ = 0.56 $\pm$ 0.23. We estimate $R_v$ = 1.97 $\pm$ 0.54 
which indicates the small size of dust particles as compared to their galactic counterparts.
A comparison of the absolute magnitude $M_B$ = -19.20 $\pm$ 0.18 with that of
other normal type Ia SNe shows that SN 2003hx is comparable in brightness to SN 1999aw.
SN 2003hx matches well with the rest of the SNe Ia sample and shows a typical linear behaviour
in $M_B$ vs $\Delta m_{15}(B)$ relation.
The bolometric light curve indicates the decay of total luminosity of the SNe. The peak
bolometric luminosity logL = 43.01 yields a value of $^{56}$Ni mass ejected to be
0.66 $M_\odot$. Comparing the Nickel masses ejected for different SNe Table \ref{comparison_sn}
we see that the ejected mass of $^{56}$Ni for SN 2003hx is slightly the higher side. The spectral 
evolution indicates SN 2003hx to be a normal type Ia event.
%
%
\section*{Acknowledgement}
We thank the observing staff in HCT for the observations.
We are thankful to an anonymous referee for helpful comments and suggestions.
This research has made use of data obtained through the High Energy Astrophysics Science
Archive Research Center Online Service, provided by the NASA/Goddard Space Flight Center.
IRAF is distributed by the National Optical Astronomy Observatories, which are operated
by the Association of Universities for Research in Astronomy, Inc., under contract to the
National Science Foundation.

\end{document}